\documentclass[12pt,preprint]{aastex}

\newcommand{\HMS}[3]{$#1^{\mathrm{h}}#2^{\mathrm{m}}#3^{\mathrm{s}}$}
\newcommand{\DMS}[3]{$#1^\circ #2' #3''$}
\newcommand{\g}{$\gamma$}

\newcommand{\etacar}{$\eta$~Carina}
\newcommand{\order}{$\mathcal{O}$}

\shorttitle{Particle acceleration in \etacar's blast wave}
\shortauthors{Ohm, Hinton \& Domainko}

\begin{document}

\title{Particle acceleration in the expanding blast wave of \etacar's 
  Great Eruption of 1843}

\author{S. Ohm\altaffilmark{1}}
\affil{Max-Planck-Institut f\"ur Kernphysik, P.O. Box 103980, D 69029
  Heidelberg, Germany}
\email{stefan.ohm@mpi-hd.mpg.de}

\and

\author{J.~A. Hinton\altaffilmark{2}}
\affil{Department of Physics \& Astronomy, University of Leicester, University Road, 
Leicester LE1 7RH, United Kingdom}
\email{jim.hinton@leicester.ac.uk}

\and

\author{W. Domainko\altaffilmark{1}}
\affil{Max-Planck-Institut f\"ur Kernphysik, P.O. Box 103980, D 69029
  Heidelberg, Germany}
\email{wilfried.domainko@mpi-hd.mpg.de}

\begin{abstract}
Non-thermal hard X-ray and high-energy (HE; $1~{\rm MeV}\leq {\rm E} \leq100~{\rm GeV}$)
\g-ray emission in the direction of \etacar\ has been recently detected using the 
\emph{INTEGRAL, AGILE} and \emph{Fermi} satellites. So far this emission has been 
interpreted in the framework of particle acceleration in the colliding wind region 
between the two massive stars. However, the existence of a very fast moving blast 
wave which originates in the historical 1843 ``Great Eruption'' provides an 
alternative particle acceleration site in this system. Here we explore an alternate 
scenario and find that inverse Compton emission from electrons accelerated in the 
blast wave can naturally explain both the flux and spectral shape of the measured 
hard X-ray and HE \g-ray emission. This scenario is further supported by the lack 
of significant variability in the \emph{INTEGRAL} and \emph{Fermi} measured fluxes.
\end{abstract}


\section{Introduction}

For a long time \etacar\ -- one of the most peculiar objects in the Milky Way -- was 
believed to be a hypergiant luminous blue variable (LBV) star. Recent 
observations, however, suggest that it is a binary system composed of a massive 
primary ($M \geq 90M_{\odot}$) and a less massive 
secondary ($M \leq 30 M_{\odot}$) 
\citep[see e.g.][]{EtaCar:Hillier01,EtaCar:Pittard02,EtaCar:Nielsen07}. 
\etacar\ experienced a historical outburst in the 19$^{\rm th}$ century and ejected 
$\sim 12\,M_{\odot}$ \footnote{\citet{EtaCar:Gomez10} conclude from submillimetre 
observations that more than $\sim 40\,M_{\odot}$ could have been ejected in the 
giant outburst. We restrict ourselves to the more conservative estimate of the 
kinetic energy in the expanding material.} of gas which moves outwards at an 
average speed of $\sim650$~km~s$^{-1}$ \citep{EtaCar:Smith03}. This material formed 
the {\it Homunculus Nebula} which is expanding with a kinetic energy of 
$\approx (4-10)\times10^{49}$~erg -- roughly 10\% of the energy released by a 
supernova explosion. 

The optical \citep{EtaCar:Damineli96,EtaCar:Damineli00} and IR 
\citep{EtaCar:Whitelock94,EtaCar:Whitelock04} lightcurves suggest a binary period of 
5.5 years, moreover, the analysis of the X-ray lightcurves points to an highly 
eccentric orbit of $e \sim 0.9$ \citep{EtaCar:Corcoran01,EtaCar:Okazaki08}. \etacar\ 
is a colliding wind binary (CWB) with powerful winds produced by both members of the 
system. The primary has a very high mass-loss rate of 
$\dot{M}_1 \geq 5\times 10^{-4}~M_{\odot}~{\rm yr}^{-1}$ 
\citep{EtaCar:Hillier01, EtaCar:Parkin09} and a terminal wind velocity of 
$v_1 \approx (500 - 700)~\mathrm{km~s^{-1}}$, whereas the secondary has 
a fast low-density wind with $v_2 \approx 3000~\mathrm{km~s^{-1}}$ and 
$\dot{M_2} \approx 1.0\times 10^{-5}~M_{\odot}~{\rm yr}^{-1}$ \citep{EtaCar:Pittard02}. 
When supersonic expanding winds collide, they form a wind-wind interaction zone 
where charged particles can in principle be accelerated to high energies via the 
diffusive shock acceleration (DSA) process 
\citep[see e.g.][]{CWB:Eichler,CWB:Reimer06}. Observationally, particle 
acceleration in the wind-collision region of massive binary systems such as \etacar\ 
is suggested by the detection of non-thermal radio emission in several objects 
\citep[see][for a review]{CWB:Review}. This indicates the existence of high-energy 
(HE; $1~{\rm MeV}\leq {\rm E} \leq100~{\rm GeV}$) electrons in these systems. 

\begin{figure}
\epsscale{.95}
\plotone{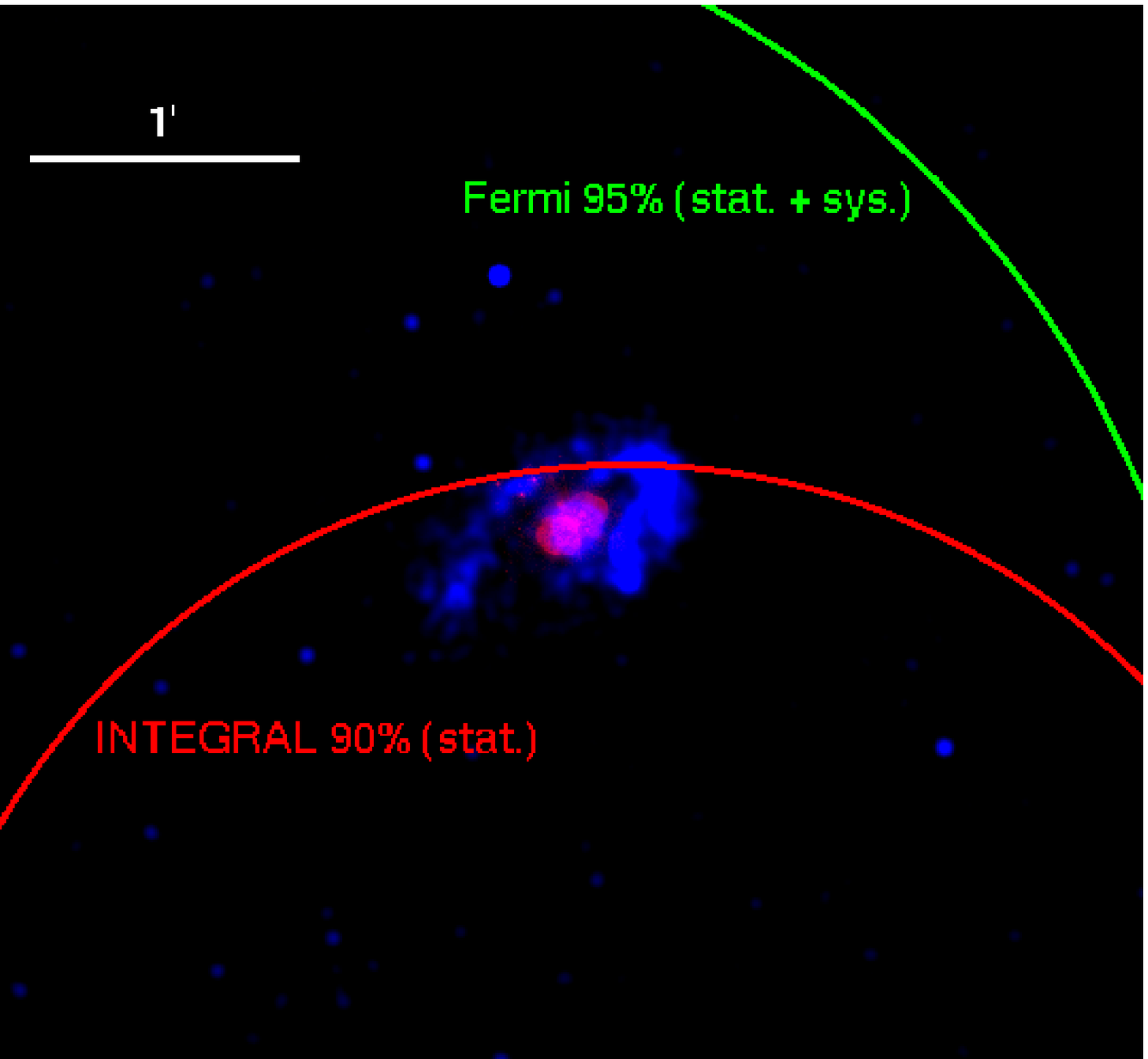}
\caption{Color-coded image of \etacar\ with X-ray emission ($0.5-11$~keV) as 
  measured by \emph{Chandra} (ObsID: 1249) in linear scale shown in blue and optical 
  emission as measured with the HST in log scale in red. Overlaid in red and green 
  are the $2.8'$ \emph{INTEGRAL} (90\% stat., \citet{EtaCar:Integral}) and $3.6'$ 
  LAT (95\% stat. + sys., \citet{Fermi:1yr}) positional uncertainties of the hard 
  X-ray and HE \g-ray source positions, respectively. The positional uncertainty  
  the \emph{AGILE} source encompasses the entire field. The best 
  fit positions and corresponding errors are summarised in Table~\ref{tab:etacar_pos}.
  \label{fig:etacar_sketch}}
\end{figure}

There is also evidence for the presence of relativistic particles in \etacar\ with 
non-thermal X-ray emission recently reported by the 
\emph{INTEGRAL} collaboration. The authors obtain an average luminosity of 
$7\times10^{33}~\mathrm{ergs~s^{-1}}$ in the hard X-ray band $(22-100)$~keV 
\citep{EtaCar:Integral}. In the HE domain, a source spatially coincident 
with the \etacar\ position, and possibly variable, was reported by the \emph{AGILE} 
collaboration \citep{EtaCar:Agile}. A steady source in the same direction 
(including systematical and statistical errors, see Fig.~\ref{fig:etacar_sketch} 
and Tab.~\ref{tab:etacar_pos}) was confirmed using the LAT instrument aboard the 
\emph{Fermi} satellite \citep{Fermi:BSL,Fermi:1yr}. The luminosity in the HE domain 
between 100~MeV and 100~GeV as obtained from the energy flux given in 
\citet{Fermi:1yr} for a source distance of 2.3~kpc is 
$1.6\times10^{35}~\mathrm{ergs~s^{-1}}$, a factor of 20 higher than the 
\emph{INTEGRAL} luminosity. 

\begin{table*}
  \begin{center}
    \begin{tabular}{l c c c}
      \hline\hline
      Instrument 
      & Energy range
      & Position (J2000) 
      & Error [$'$]
      \\      
      \hline
      \emph{INTEGRAL} & 22~keV -- 100~keV & \HMS{10}{45}{02}, \DMS{-59}{43}{38} & 2.8  \\
      \emph{Fermi} & 30~MeV -- 30~GeV & \HMS{10}{45}{13}, \DMS{-59}{42}{21} & 3.6  \\
      \emph{AGILE} & 100~MeV -- 100~GeV & \HMS{10}{44}{49}, \DMS{-59}{44}{54} & 24  \\
      \hline\hline
    \end{tabular}
  \end{center}
  \caption{Best fit positions of the sources detected in the hard X-ray and HE 
      \g-ray band. The 90\% (stat.) error on the position of the \emph{INTEGRAL} 
      source as well as the 95\% (stat. + sys.) uncertainty on the source position 
      of \emph{Fermi} and \emph{AGILE} are given.}
  \label{tab:etacar_pos}
\end{table*}

Given the large population of HE \g-ray sources detected by \emph{Fermi}, with a 
significant fraction associated with pulsars, the chance probability of source 
confusion with a pulsar or another HE \g-ray source must be examined. 
Within a circle of $5^\circ$ radius, 19 ATNF pulsars with a spin-down luminosity 
$\dot{E} > 3\times 10^{33}$~erg/s \citep{ATNF:Cat} -- the sensitivity limit of the 
\emph{Fermi} \g-ray pulsar catalog \citep{Fermi:PSRCat} -- and 12 \emph{Fermi} 
sources \citep{Fermi:1yr} have been detected. Thus, the probability to find any of 
the HE \g-ray sources or ATNF pulsars within the \emph{Fermi} error circle at the 
\etacar\ position is estimated to less than 0.2\%. Moreover, whereas pulsars 
typically have a cut-off in the energy spectrum $\leq 10$~GeV, the spectrum of the 
\emph{Fermi} source positionally coincident with \etacar\ shows significant flux 
above 20~GeV, making an association with a pulsar unlikely.

\begin{figure}
\epsscale{.95}
\plotone{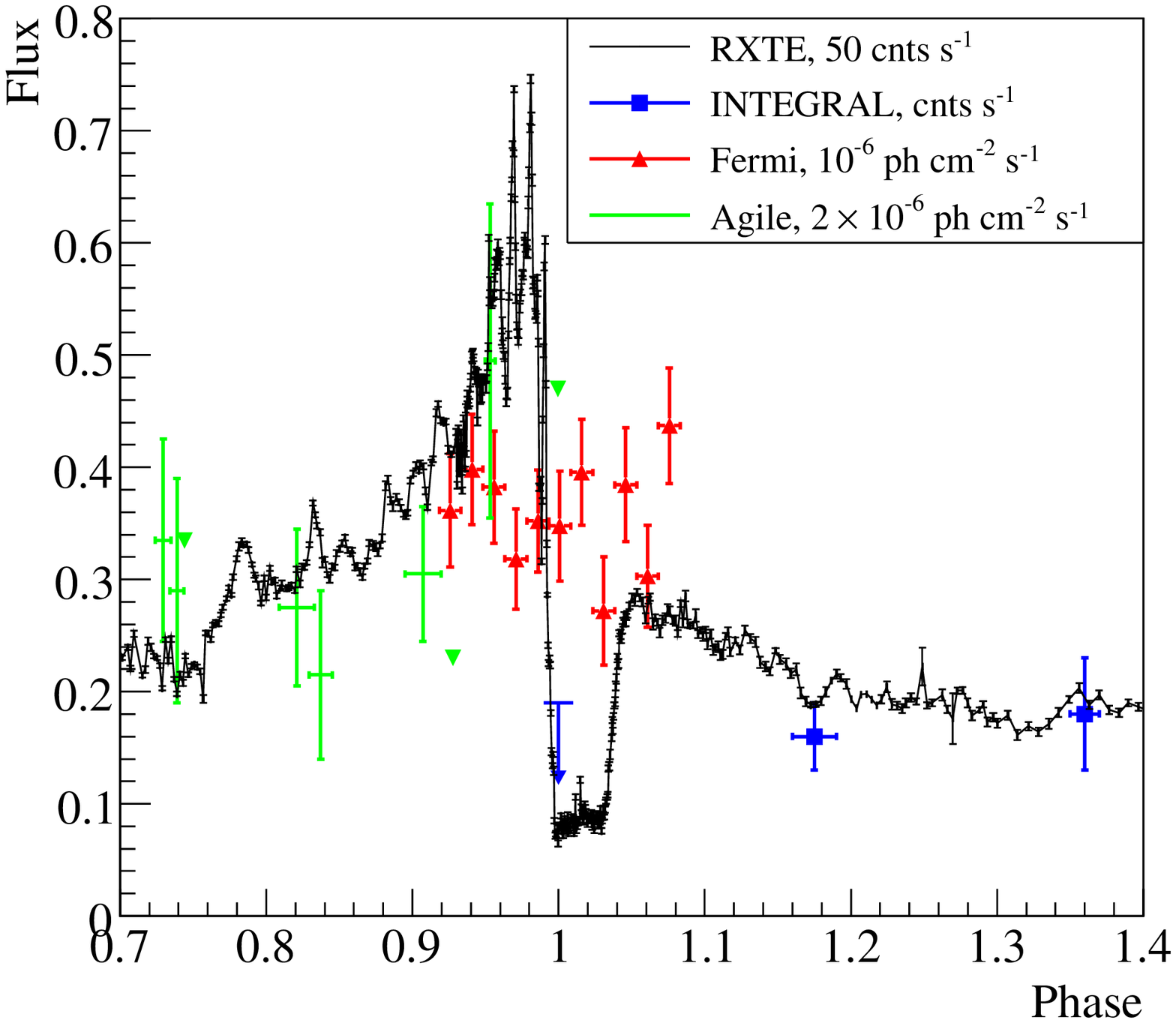}
\caption{\etacar\ lightcurve as a function of orbital phase as measured by RXTE 
  (black), \emph{INTEGRAL} (blue), \emph{AGILE} (green) and the \emph{Fermi}-LAT 
  instrument (red). \label{fig:etacar_lc}}
\end{figure}

The lightcurves from all mentioned instruments, together with lower-energy X-ray 
data from RXTE, which probes the colliding wind region (CWR), are shown in 
Fig.~\ref{fig:etacar_lc}. Note, that the flare reported by \emph{AGILE} around phase 
$p\approx0.96$ was not confirmed using the more sensitive \emph{Fermi}-LAT. 
Although \emph{INTEGRAL} detected no significant hard X-ray emission during 
periastron passage, the upper limit is consistent with steady emission.
The absence of significant variability in the 50~keV and GeV regime as shown  
in Fig.~\ref{fig:etacar_lc} is surprising in a CWB picture, especially during 
periastron passage where a collapse of the CWR is expected and hence no particle 
acceleration should occur.

The detections of hard X-ray and HE \g-ray emission in \etacar\ have so far been 
interpreted in the framework of particle acceleration in the colliding winds of the 
two massive stars. However, the CWR of \etacar\ is not the only place within the 
\emph{Fermi}, \emph{AGILE} and \emph{INTEGRAL} error circles, where particles can be 
accelerated via DSA to high energies. Recent observations show that \etacar\ is 
surrounded at a distance of $\sim$0.25~pc by a very fast moving blast wave.
The giant outburst of 1843 (also known as the ``Great Eruption'') produced material 
which is now moving ahead of the {\it Homunculus Nebula} at speeds of 
$3500-6000$~km~s$^{-1}$ \citep{EtaCar:Smith08}. This blast wave currently 
overruns the ``Outer Ejecta'' a ring-like structure of material which originates 
from an ejection of mass from \etacar\ $\sim500-1000$ years ago 
\citep{EtaCar:Walborn78}.

The existence of the fast-moving material doubles the estimate of the total kinetic 
energy of the giant outburst and mimics a low-energy supernova remnant (SNR) shell 
\citep{EtaCar:Smith08} with a blast wave moving into the ISM with velocities 
comparable to the historical supernovae RCW~86 \citep{Vink:RCW8606} and SN~1006 
\citep{SN1006:Vink05}. Fig.~\ref{fig:etacar_sketch} shows the X-ray emission as 
measured with \emph{Chandra} and optical emission measured with the 
\emph{Hubble Space Telescope} (HST). The 95\% (stat. + sys.) and 90\% (stat.) error 
circles of \emph{Fermi}-LAT and \emph{INTEGRAL}, respectively, are overlaid. They 
enclose the \etacar\ position. In this work the possibility of particle acceleration 
in the expanding blast wave of 1843's Great Eruption is examined and leptonic and 
hadronic origins of the observed \g-ray emission are investigated in detail. 

\section{Particle acceleration in the blast wave}

SNRs start their evolution by a free-expansion phase of the ejecta and enter the 
pressure-driven Sedov phase, once the mass of displaced material roughly equals the 
ejecta mass. The transition between the two phases in the remnant surrounding 
\etacar\ is expected to occur after $290~(500)~n_H^{-1/3}~{\rm years}$ depending on 
the ambient density $n_H$, for a blast wave speed of 6000~(3500)~km/s, 
respectively. The density of the medium surrounding the blast wave can be estimated 
by adopting two steady stellar winds which filled the volume around \etacar\ prior 
to the Great Eruption. For typical wind speeds of 500~km/s and a typical mass-loss 
rate of $5-10 \times 10^{-4}$ M$_{\odot}$ yr$^{-1}$ 
\citep{EtaCar:Hillier01,EtaCar:Parkin09} the density of the circumstellar medium 
would be of the order of 50~cm$^{-3}$ at a distance of 0.25~pc. Hence, the 
blast wave is currently in the transition between free-expansion and Sedov phase, implying 
that efficient particle acceleration is possible. 

The distance of 0.25~pc of the blast wave from \etacar\ used here is motivated 
by the conclusion of \citet{EtaCar:Smith08} that parts of the Outer Ejecta are 
currently being overrun by the blast wave and by X-ray observations, where 
\citet{EtaCar:Seward01} used a ring with the same radius to describe the soft X-ray 
shell coincident with the Outer Ejecta. On the other hand, for a steady shock 
speed of 3500~km/s and an age of 167 years, the expected distance of the blast wave 
to \etacar\ is 0.6~pc. A wind solution of the fast 
moving material, which would result in an increasing velocity with time, could 
explain such a difference. To account for the different distance estimates, 
the spectral energy distributions (SEDs) for both values are 
shown in Section~\ref{sec:SED}.

Relativistic electrons and protons will gain energy when crossing the shock 
front of \etacar's blast wave. In the limit of a strong shock, the particle 
crossing time $\Delta t$ is determined by the diffusion coefficient $\kappa$, 
expressed in terms of the Bohm diffusion coefficient:
\begin{equation}
\kappa = \eta\kappa_{\rm Bohm} = 
10^{23}~\eta~{\rm cm}^2~{\rm s}^{-1}~\left(\frac{E_{\rm GeV}}{B_{\mu \rm G}}\right)\ ,
\end{equation}
the shock speed $v_{s}$, the particle energy $E_{\rm GeV}$ and the magnetic field in 
the shock region $B_{\rm \mu G}$ as:
\begin{equation}
\Delta t = \frac{5\eta\kappa_{\rm Bohm}}{cv_{s}} 
= 1.75\times10^5~{\rm s}~\left(\frac{E_{\rm GeV}}{v_{{s,10^3~\rm km~s}^{-1}}B_{\mu \rm G}}\right)\ .
\end{equation}
The energy gain per shock crossing is given by $\Delta E/E = v_{s}/c$ 
and, hence, the acceleration timescale 
$\tau_{\rm acc} = E/(\Delta E/\Delta t) \approx \eta\kappa / v_{s}^2$: 
\begin{equation}
\tau_{\rm acc} \approx 5\times10^7~{\rm s}~\eta~v_{{s,10^3~\rm km~s}^{-1}}^{-2}\left(\frac{E_{\rm GeV}}{B_{\mu \rm G}}\right)\ .
\end{equation}
No magnetic field measurements are available in the region of the blast wave. 
However, for the {\it Homunculus Nebula} different estimates exist and are used in 
the following as a rough guide for the magnetic field strength in the region of 
interest.
\citet{EtaCar:Bfield} derived from polarization measurements, which were based 
on dust grain alignment, magnetic field strengths of \order($\mu$G--$m$G) in the 
{\it Homunculus Nebula}, depending on the underlying process. However, the dust 
shell of the \emph{Homunculus} appears to be neutral and is likely composed of 
silicates and Fe rather than typical ISM dust grains \citep{EtaCar:Gail05}. Hence, 
for the purpose of this work, a lower magnetic field strength of $10\, \mu$G is 
assumed. In the Bohm limit and without energy losses, it would take about a year to 
accelerate an electron or proton to an energy of 100~GeV in a $10\,\mu G$ field 
given a shock speed of 3500~km~s$^{-1}$. This acceleration timescale is short compared
to the time since the giant outburst (167 years) and implies that 
charged particles can potentially be accelerated to TeV energies.


\section{Origin of the \g-ray emission}

Protons and heavier nuclei accelerated in the expanding blast wave will interact 
with the ambient medium, producing $\pi^0$-decay \g\ rays above $\sim$300 MeV. The 
timescale for inelastic proton-proton interactions is somewhat energy-dependent but 
can be approximated as \citep{Felix:Book}:
\begin{equation}
\tau_{pp} = 3\times 10^7~n_H^{-1}~{\rm yrs}\ ,
\end{equation}
where $n_H$ is the density of hydrogen atoms per cm$^{-3}$.
From a veil of [O III] emitting material surrounding \etacar\ and its ejecta  
\citet{EtaCar:Smith05} estimate a mean density of 400 particles per cm$^{-3}$ and 
\citet{EtaCar:Smith08} finds 500 cm$^{-3}$ in this region, implying 
$\tau_{pp} = 6-8\times10^4$ years. This timescale is much longer 
than the age of the remnant and hence the p-p channel is radiatively inefficient. 
To explain the observed level of \g-ray emission with this density of target 
material an energy of about 
$L_{\gamma} \epsilon_{\gamma} \tau_{pp} \sim 3 \times 10^{48}$~erg in hadronic cosmic 
rays ($\epsilon_{\gamma}\sim0.1$ is the fraction of the proton energy released in 
\g\ rays in a typical interaction) would be required. This is roughly 5\% of 
the kinetic energy of the blast wave, quite plausible within the framework of DSA. 
However, it is hard to explain the \emph{Fermi} emission below 300 MeV, and 
especially the \emph{INTEGRAL} emission, in this scenario. 
Furthermore the required maximum energy of protons as indicated by the 
curvature in the \emph{Fermi} spectrum is well below the expected maximum 
acceleration energy associated with either the age or size of the system. 

Contrary to hadronic cosmic rays, accelerated electrons will encounter significant 
energy losses during the acceleration process. They will predominantly lose energy 
via inverse Compton (IC) scattering on radiation fields present in the acceleration 
region, via synchrotron radiation in magnetic fields, and by Bremsstrahlung in 
interactions with ambient material. 
The IC cooling time in the presence of 
a black-body target radiation field with energy density $U_{\rm rad,~eV~cm^{-3}}$ for an 
electron of energy $E_{\rm GeV}$ is \citep{Moderski05,HESS:Review}:
\begin{equation}
\tau_{\rm loss} \approx 3.1\times10^8~{\rm yrs}~U_{\rm rad,~eV~cm^{-3}}^{-1}~E_{\rm GeV}^{-1}~f_{\rm KN}^{-1}\ .
\end{equation}
Assuming that the radiation field of the whole {\it Homunculus Nebula} is 
concentrated in a point at 0.25~pc distance to the blast wave and follows a 
black-body spectrum of 260~K temperature \citep{EtaCar:Gehrz99}, the energy density 
in IR photons which 
serve as target radiation field is $U_{\rm rad} = 8\times10^4~{\rm eV~cm}^{-3}$,
several orders of magnitude higher than the density of the cosmic microwave 
background ($0.27~{\rm eV~cm}^{-3}$) or typical ISM photon fields. For infrared 
target photons the Klein-Nishina (KN) suppression of the IC cross-section for 
electron energies $E \leq 100$~GeV is a less than 20\% effect and is not considered 
in the following. The relevant IC cooling time for electrons in the blast wave region 
is therefore $\tau_{\rm loss} \approx 4000~{\rm yrs}~E_{\rm GeV}^{-1}$. In a 
$10\,\mu{\rm G}$ field, the synchrotron-loss timescale for 100~GeV electrons is of 
\order($10^6$) years: IC losses will dominate unless $B>2{\rm mG}$.
In dense media, Bremsstrahlung may also contribute significantly to the energy-loss 
rate of electrons. The estimated mean density of $400-500$ particles per cm$^{-3}$ 
implies a typical Bremsstrahlung loss time of 
$\tau_{\rm Brems} \approx 8-11\times 10^4$~years. Therefore, at $>$ GeV energies, 
IC emission is very likely to be the dominant emission process for electrons in this 
system.

The maximum electron energy is determined (for cooling-limited acceleration) by the 
balance between radiative energy losses and the rate of energy gain via DSA, when 
$\tau_{\rm loss} = \tau_{\rm acc}$:
\begin{equation}
E_{e,\rm max} \approx 50~{\rm GeV}~v_{{s,10^3~\rm km~s}^{-1}}\sqrt{\frac{B_{\mu \rm G}}{\eta}}\ .
\label{eq:emax}
\end{equation}
This maximum energy for the measured blast wave speed is consistent with the HE 
\g-ray emission found by \emph{Fermi} 
for moderate magnetic fields of $10\,\mu{\rm G}$ and diffusion close to 
the Bohm limit. Moreover, the energy-loss timescale is much shorter than the age of 
the remnant, implying that acceleration of electrons is indeed cooling limited in 
this system. In this scenario most of the energy injected into the acceleration 
process emerges as IC radiation in the GeV domain. For such an efficient process the 
required energy input in electrons is modest: within an order of magnitude of 
$L_{\gamma} t_{\rm sys} \sim 10^{45}$~erg. IC emission is expected to dominate over 
p-p emission in this system as long as $E_{e}>10^{-3} E_{\rm CR}$.

\citet{Vannoni09} performed numerical calculations where they studied acceleration 
and radiation of electrons in an radiation-dominated environment such as the one 
investigated in this work. Compared to the magnetic field of $10\,{\mu \rm G}$ used 
here, the authors obtain for magnetic fields of $100\,{\mu \rm G}$ maximum particle 
energies in the TeV regime, where KN effects become very significant.

\section{Spectral Energy Distribution}\label{sec:SED}

\begin{figure}
\epsscale{1}
\plotone{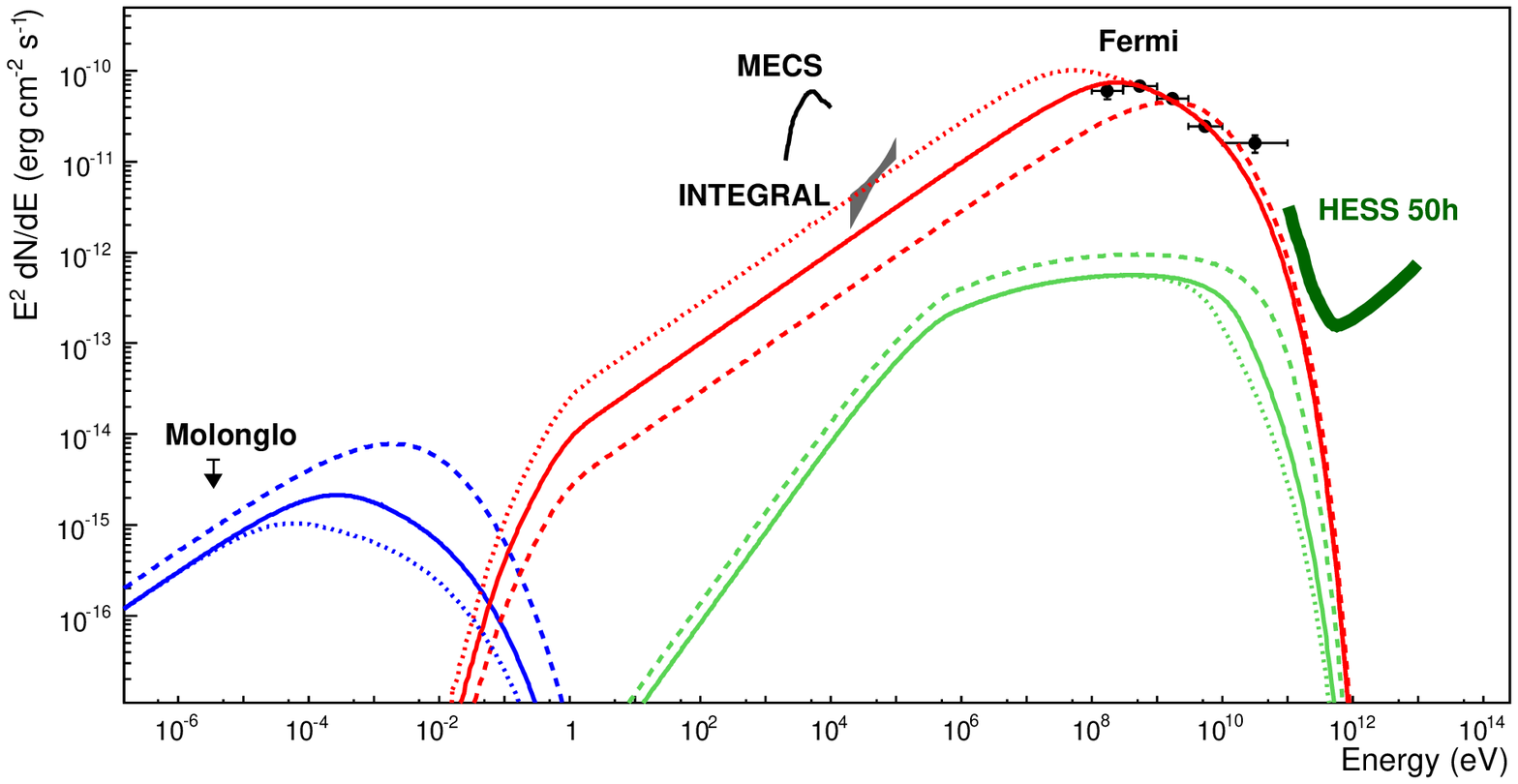}
\caption{Measured and calculated SED for the region within 
a few arcminutes of \etacar, showing the radio flux measured by Molonglo 
\citep{Molonglo} as an upper limit on the synchtrotron flux, thermal X-ray emission 
as measured by MECS, which probes the wind-wind interaction region, 
\citep{MECS1,MECS2}, the \emph{INTEGRAL} measurement from 
\citet{EtaCar:Integral} and the \emph{Fermi}-LAT detection \citep{Fermi:1yr}. 
The sensitivity of the H.E.S.S. instrument at TeV energies is shown for comparison. 
The curves show a single zone time-dependent model for continous injection of 
electrons over the 167 year history of the system and a maximum acceleration energy 
of 110\,GeV. A radiation density of $8\times10^4~{\rm eV~cm}^{-3}$, appropriate for 
a region at 0.25 pc from \etacar, is used for the solid curves, decreased to 
$1.4\times10^4~{\rm eV~cm}^{-3}$, corresponding to 0.6~pc distance, for the dashed 
curves and increased to $2.2\times10^5~{\rm eV~cm}^{-3}$, corresponding to 0.15~pc 
distance, for the dotted curves. A magnetic field strength of $10\,\mu$G 
is assumed. The Bremsstrahlung component for each model is depicted in green.}
  \label{fig:sed}
\end{figure}

Figure~\ref{fig:sed} shows the measured SED of the $1'$ region around \etacar\ 
together with calculated synchtrotron and IC emission for the leptonic scenario 
outlined above (solid lines). The calculation is a single-zone time-dependent 
numerical model as described in \citet{HintonAharonian2007}. 

Electrons accelerated via the DSA process are expected to
follow a power-law in energy $dN/dE_e\sim E_e^{-\Gamma_e}$ with index
$\Gamma_e\approx 2$ and produce an IC (and synchrotron)
\g-ray spectrum with spectral index $\Gamma_\gamma = -(\Gamma_e+1) / 2$. 
For continous injection IC cooling in the Thomson regime results in a spectral break 
with $\Delta\Gamma_\gamma = 1/2$. In the case of the Great Eruption 167
years ago, the IC break energy is expected to lie at $\approx
0.4$~GeV  for a blast wave distance of 0.25\,pc. As Figure~\ref{fig:sed} 
shows, for this distance the observed break between the \emph{INTEGRAL} and 
\emph{Fermi} spectra occurs somewhat below this energy, close to the 
expectation for the higher radiation field corresponding to a smaller distance of 
0.15\,pc. This value is broadedly consistent with expectations since the relevant 
radiation field is the average value encountered over the history of the object. 
Furthermore, the system is far from radially symmetric and parts of the shell lie 
at significantly smaller than average distances to the \emph{Homunculus}. The 
\emph{Fermi} data indicate a high energy cut-off at an electron energy of 
$\approx100$ GeV, very close to the expected cut-off for IC-cooling-limited 
acceleration.

The dashed and dotted-lines in Figure~\ref{fig:sed} demonstrate the effect of 
adjusting the main parameter of the model. For the dashed curves the 
radiation field is decreased to that expected at 0.6~pc (rather than 0.25~pc) 
from \etacar. In this case an increase in energy in electrons by a factor of 
2.3 would be required to match the \emph{Fermi} flux. For the dotted curves the 
radiation field is increased to that expected at 0.15~pc from \etacar.
All curves were produced assuming a magnetic field of $10\,\mu$G. Magnetic 
fields higher than $\sim20\,\mu$G are excluded in this scenario by the limit on 
non-thermal radio emission from Molonglo. For a blast wave 
distance of 0.6~pc, the shape of the resulting model spectrum reproduces the data in 
the HE domain less well. A model with a realistic time and space dependent radiation field
would be expected to fall in the region defined by the three curves shown.

\section{Summary}

It appears that the properties of the high-energy non-thermal emission of \etacar\ 
are consistent with an origin at the high-velocity blast wave of the Great Eruption 
of 1843. As Figure~\ref{fig:sed} demonstrates, the position of the spectral break 
between the \emph{Fermi} and \emph{INTEGRAL} domains, and the cut-off energy measured 
using \emph{Fermi}, can be explained as a cooling break and an acceleration limit, 
respectively. 

A prediction of this model is an extended emission region and hence 
non-variability of the signal. The non-variability of the source in the \g-ray 
regime is supported by the \emph{Fermi} observations. While \emph{Fermi} and 
\emph{INTEGRAL} do not provide sufficient angular resolution to resolve the blast 
wave (see Fig. \ref{fig:etacar_sketch}), high resolution radio observations will be 
able to resolve non-thermal radio emission from this acceleration site. Detection of a 
radio shell at the location of the shock would support the proposed scenario. 

The time elapsed since the ejection of the very fast material by \etacar\ is 
comparable to the age of the youngest Galactic SNR G\,1.9+0.3 and between the ages 
of the remnants of SN\,1987A and Cas\,A. It may provide important insights into 
particle acceleration and production of non-thermal radiation during the early 
stages of remnant evolution where only very few objects can be studied. 

\acknowledgments
We would like to thank the anonymous referee for very helpful and informative 
comments which improved the quality of the paper. We would also like to thank Felix 
Aharonian, Werner Hofmann, Olaf Reimer and Klaus Reitberger for their input and 
suggestions. Furthermore, the support of the authors host institutions, and 
additionally support from the German Ministry of Education and Research (BMBF) is 
acknowledged.


\clearpage

\end{document}